\documentclass[twocolumn,3p]{elsarticle} 
\usepackage{lineno} 
\usepackage{mathtools}
\usepackage{graphicx} 

\journal{Nucl. Instr. and Meth. A}

\begin{document} 

\begin{frontmatter}

\title{An LED pulser for measuring photomultiplier linearity} 

\author[1]{M. Friend\corref{cor}}
\ead{mfriend@andrew.cmu.edu}
\author[1]{G. B. Franklin} 
\author[1]{B. Quinn}

\address[1]{Carnegie Mellon University, Department of Physics, 5000 Forbes Ave, Pittsburgh, PA 15213, USA}

\cortext[cor]{Corresponding author. Tel: +1-412-268-6949, fax: +1-412-681-0648}

\begin{abstract}

A light-emitting diode (LED) pulser for testing the low-rate response of a
photomultiplier tube (PMT) to scintillator-like pulses has been designed,
developed, and implemented.  This pulser is intended to simulate 80 ns full
width at half maximum photon pulses over the dynamic range of the PMT, in order
to precisely determine PMT linearity.  This particular design has the advantage
that, unlike many LED test rigs, it does not require the use of multiple
calibrated LEDs, making it insensitive to LED gain drifts.  Instead, a
finite-difference measurement is made using two LEDs which need not be
calibrated with respect to one another.  These measurements give a better than
1\% mapping of the response function, allowing for the testing and development
of particularly linear PMT bases.

\end{abstract}


\end{frontmatter}

\section{Introduction} 

An LED pulser has been designed to accurately map out the response of a PMT (in
this case, an RCA 8575) to pulses with a full width at half maximum (FWHM) of 80
ns.  For this application, the output pulse height of interest was up to about
2.5 V.  The pulse range used was meant to simulate the response of a Ce-doped
\(\text{Gd}_2\text{SiO}_5\) (GSO) crystal to photons ranging from 1 to 600 MeV,
although pulses of different widths and heights may also be generated using this
device.  The particular application of interest, a Compton backscattering
polarimeter read out by charge integration \cite{Friend:ComptMeas11}, required
excellent and well-understood linearity in PMT response.

The response function of a PMT is defined as the signal output of the PMT given
some light input.  If the integrated response, \(f(x)\), to a light flash of
integrated brightness \(x\) were perfectly linear, then the increase in response
resulting from an additional simultaneous flash, \(\delta\), would be constant;
i.e.\ 
\begin{equation} \label{eq:findif} y(x) \equiv f(x + \delta) - f(x),
\end{equation}  
the finite difference function of the response, would be independent of \(x\).
Saturation of the PMT response would manifest itself by a smaller increment in
response due to the fixed signal \(\delta\) as it is added onto progressively
larger signals \(x\) (i.e.\ \(y(x)\) would decrease with increasing \(x\)).
Conversely, a PMT base design which over-compensates for saturation would give
progressively larger responses to a fixed signal added onto progressively larger
signals (i.e.\ \(y(x)\) would increase with increasing \(x\)).  Any variation of
\(y(x)\) as \(x\) is varied is thus a sensitive measure of non-linearity.

This measurement is achieved by flashing two LEDs, one of constant low
brightness, called here the ``delta'' LED, which contributes a flash of
integrated brightness \(\delta\), and another of variable brightness, called the
``variable'' LED, which contributes a flash of integrated brightness \(x\).  A
finite difference measurement is then made by flashing both LEDs concurrently,
to measure \(f(x+\delta)\), and then subtracting \(f(x)\), found by flashing
just the variable LED.  Because this setup measures the response of a PMT to the
difference between a changing LED and a constant one, it is insensitive to
calibration between the LEDs (unlike, e.g., Ref. \cite{Vicic:PMTResponse03}).
It is, however, critically important that the two LEDs be independent -- there
cannot be cross-talk between the LEDs.  Also, one LED should have a low constant
amplitude and the other must be varied over the dynamic range of interest.

\section{Pulser Setup} 

Two LEDs are positioned within a light-tight PMT enclosure such that they shine
diffusely on the PMT face by reflection.

\begin{figure}[h] \begin{center}
\includegraphics[bb=140 485 295 650,clip=true]{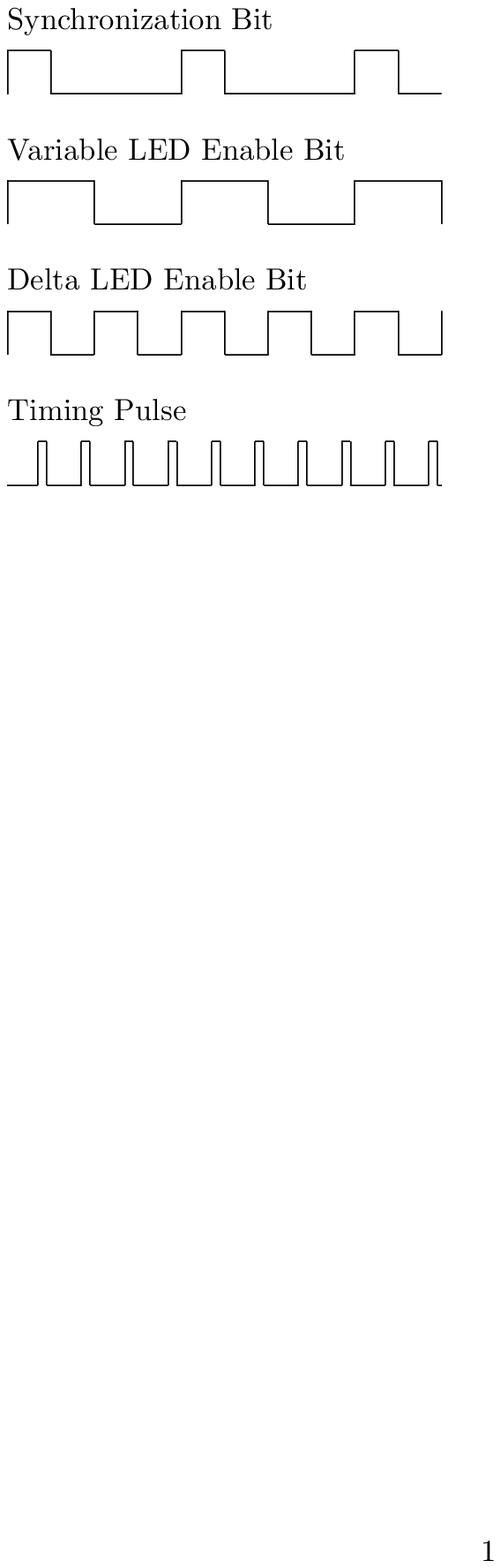} 
\caption[Timing structure for pulser bits]{Timing structure for pulser bits.
\label{fig:timing}} \end{center} \end{figure}

The LED pulser runs with a timing sequence of: (1) both LEDs flash, (2) variable
LED flashes, (3) delta LED flashes, (4) both LEDs off (shown in Fig.\
\ref{fig:timing}).  The PMT signal is read out by an Analog to Digital Converter
(ADC) at each step of the sequence.  The sequence is repeated multiple times,
then a new pulse amplitude, set by a computer controlled Digital to Analog
Converter (DAC), is chosen for the variable LED.  The variable LED setting is
repeatedly cycled through the desired range of pulse amplitudes until the
desired statistical accuracy for each pulse amplitude data point is obtained.
The both-LEDs-off step can be used for pedestal monitoring.  The delta-only step
is used only as a cross-check and is not required.  The both-LEDs-flash and
variable-LED-flashes steps are the two used for the actual response-function
measurement.  Therefore, although four steps were used for this measurement, the
sequence could be shortened.  

The delta LED is flashed at a constant, relatively low brightness.  The variable
LED is flashed at a wide range of selected brightnesses, and the brightness
setting was changed every 1/30 s.  The device was driven by a 1 kHz clock pulse,
so that the full four-step sequence ran at 250 Hz.  The delta LED was chosen to
have a brightness of roughly 1/10 the range of the variable LED, but no precise
cross-calibration of the two LEDs is required.

For readout in this particular application, the PMT pulse shape was continuously
sampled at 200 MHz by a Struck SIS3320 Flash ADC (FADC), and the clock times of
the kHz LED trigger pulses were recorded in a CAEN V830 latching scaler.  A 1000
ns sampling period was read out from the FADC memory for each clock time stored,
including any portion of the pulse stored from before the trigger. The pulse was
then integrated numerically \cite{Friend:ComptMeas11}.

\section{Pulser Design} 

The LEDs used in this pulser setup are Nichia blue, 470 nm, NSPB500S LEDs, as
recommended by Vi\'{c}i\'{c} et.\ al.\ \cite{Vicic:PMTResponse03}.  These LEDs
have the advantage that they will emit a smooth and fast scintillator-like pulse
when one side is biased with a fast pulse.  The positive leg of the LED is fed
this fast TTL pulse from a specially designed LED drive circuit, described
below, while the negative leg is given a variable DC voltage between 0 and +5 V
(set under computer-control through a DAC implemented on a VME Timing Board
\cite{Miller:th01}), which determines the brightness of the resulting flash.

The LED pulser is based on the 74AC00 NAND gate.  The fast and stable pulses
used to achieve an 80 ns FWHM LED pulse take advantage of the high current
Advanced CMOS Logic (AC) electronics, as recommended by O'Hagan et.\ al.\
\cite{Ohagan:LEDsource02}.  

Controlling cross-talk requires the division of functionality into three
separate circuits: two ``driver-boards,'' one to flash each of the two LEDs, and
one ``control-board,'' which controls the driver-boards by sending enable bits
and a signal which sets the pulse timing.  As shown in Fig.\ \ref{fig:timing},
this control circuit, which is driven by an external clock, steps the enable
bits through a binary sequence and sends a timing-pulse signal to both
driver-boards after allowing a settle-time for the enable bits.  The control
board also produces a sync pulse at the beginning of each binary sequence, to
synchronize the DAQ system.  

\begin{figure}[] 
\centering
\includegraphics[bb=95 335 505 720,clip=true, width=8cm]{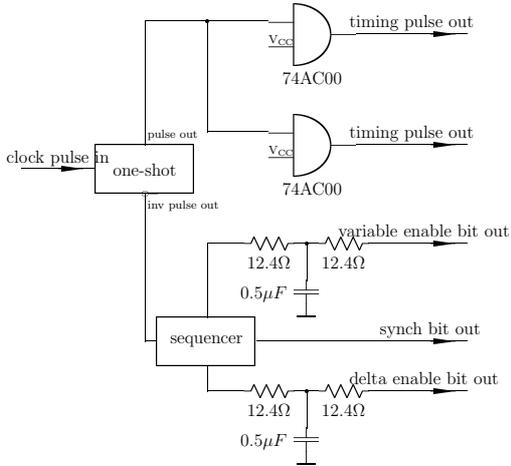} 
\caption[Control-board circuit diagram]{The circuit diagram for the control-board.  The timing output of the
sequencer is shown in Fig.\ \ref{fig:timing}. \label{fig:miniMegan}}
\end{figure}

\begin{figure}[] 
\centering
\includegraphics[bb=75 265 545 720,clip=true, width=8cm]{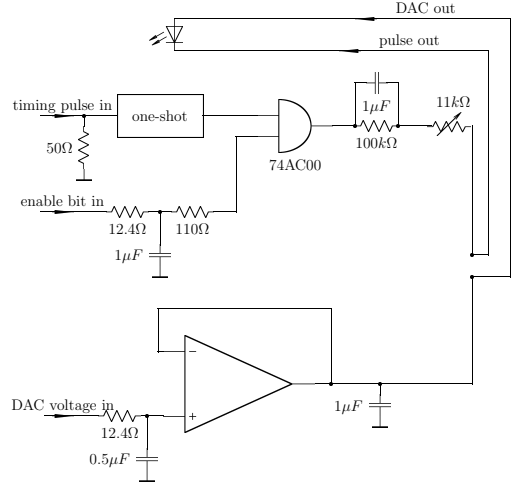} 
\caption[Driver-board circuit diagram]{The circuit diagram for the driver-board.  The variable resistor, used
to tune the LED pulse amplitude, actually consists of two trimpots in series, 1
k\(\Omega\) for fine adjustments and 10 k\(\Omega\) for coarse ones. \label{fig:microMegan}}
\end{figure}

The circuit diagrams for the control-board and driver-board are shown in Figs.\
\ref{fig:miniMegan} and \ref{fig:microMegan}.  The AND gates shown are
implemented as two 74AC00 NAND gates in succession.  The one-shot used is a
DM74LS221 monostable multivibrator with a Schmitt-trigger input.  This can be
used to delay the pulse sent to each of the driver-boards, each of which also
has a one-shot with a timing width controlled by a variable resistor.  This
variable resistor allows the user to change the pulse width as desired, but
since it is important to ensure that the flashes from both LEDs are of equal
width, the pulse widths for the two driver-boards need to be matched.  It is
also important to ensure that the pulses coming from the driver-boards are
synchronous, which is easily done by looking at the PMT pulse signal relative to
the 1 kHz clock on an oscilloscope and aligning the responses to the delta-only
and variable-only pulses.

Each driver-board reshapes the timing pulse and, if the enable bit is set, sends
the TTL pulse to the positive leg of the corresponding LED.  The DC voltage sent
to the negative leg determines the brightness of the flash in response to this
pulse.  

\subsection{Cross-Talk} 

The major design issue for this pulser setup is generating two uncorrelated LED
signals, since any cross-talk between the two LEDs destroys the finite
difference linearity measurement.  Cross-talk can be easily and precisely
measured by driving both LEDs to flash, but with one LED physically removed from
the light-tight PMT enclosure or obscured by opaque tape.  Two types of
cross-talk are then measured, as either a variation in the delta LED signal as a
function of the variable LED DAC setting when the variable LED is obscured, or
as a deviation from zero in \(f(x+\delta)-f(x)\) as a function of \(f(x)\) when
the delta LED is obscured; the former type of cross-talk was not as problematic
as the latter.  A deviation of \(0.02f(x)\) in \(y(x)\) (from the expected
\(y(x)=0\) with the delta LED obscured) over the full range of \(f(x)\) was
typical of problematic cross-talk.  The elimination of this cross-talk effect is
achieved through several important design features.  

One requirement for eliminating cross-talk is putting the driver-board for each
LED into a separate shielded box.  Each box includes low-pass filters to reduce
noise transmission.  It is critical that each driver-board's behavior be
independent of whether the other LED fires.  To ensure this, the control-board
sends the same timing pulse to both driver-boards regardless of whether they are
enabled or not.  It was found to be useful to add a low pass filter to prevent
noise returning along the enable-bit line, which was telegraphing whether the
driver-board had fired its LED.  The boxes are also physically separated from
the control-board, which is in another shielded box.

The fast pulse to turn on the LED needs to be delayed until well after the
enable bits have switched.  A settling time of 50 \(\mu\)s was used. The pulses
are also re-generated using a one-shot for each of the two LEDs, so that
variation in pulse width cannot telegraph the enable bit of one LED to the other
LED.  

Separate 5 V power supplies are also used for each of the two LED driving
circuits, as well as for the main controller circuit.

The LEDs, which connect to each driving circuit via a DE-9 connector, must have
cables leading to the PMT enclosure which are short and well shielded.
Twinaxial cables were used in this application.

It was also determined that the two LEDs cannot be placed closer together than
about 8 cm, or there is cross-talk, as previously seen \cite{Camsonne:privcom}.
Since this effect is seen even when the LEDs are optically isolated from one
another, the cross-talk is apparently electro-magnetic.  After eliminating all
other forms of cross-talk, placing the two LEDs 6 cm apart, instead of 8 cm,
contributed a clear deviation of \(0.0013f(x)\) in \(y(x)\), with the delta LED
obscured.

\section{PMT Response Function} 

\begin{figure}[] \caption[Finite difference response curve]{Finite difference response curve for pulses
corresponding to up to 600 MeV photons in GSO.  The curve is scaled by a single
factor vertically and horizontally to give a maximum of \({\sim} 1\) on the
horizontal axis.  The solid line is a fit to the response curve,
\(F_n(x+\delta)-F_n(x)\), where \(n=16\). Note that the zero is suppressed on
the vertical axis. \label{fig:d2nfindiff}}
\includegraphics[width=8.5cm]{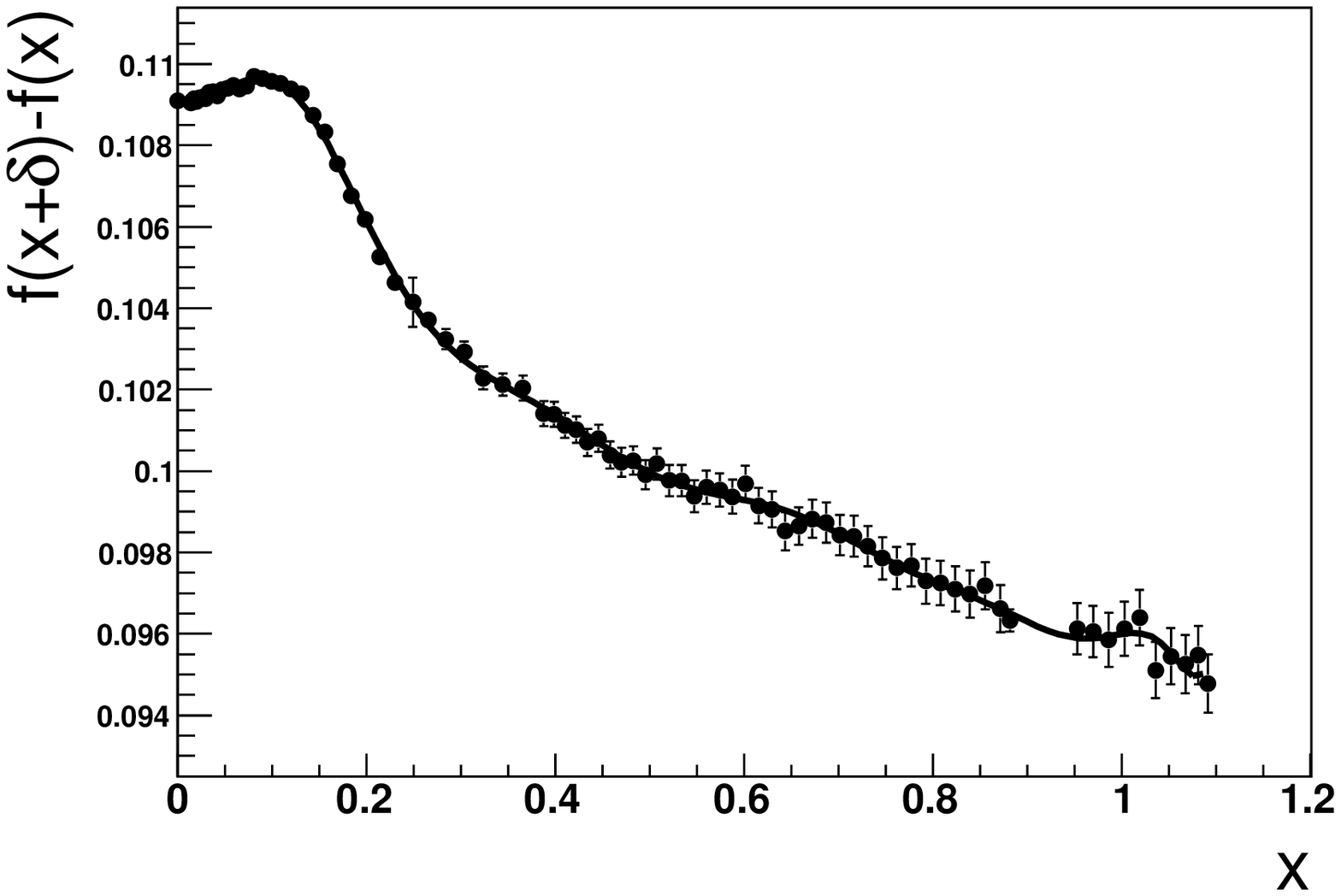} \end{figure}

\begin{figure}[] \caption[Finite difference response curve]{Finite difference response curve for pulses
corresponding to up to 200 MeV photons in GSO.  The curve is scaled by a single
factor vertically and horizontally to give a maximum of \({\sim} 1\) on the
horizontal axis.  The solid line is a fit to the response curve,
\(F_n(x+\delta)-F_n(x)\), where \(n=6\).  This response curve was generated
using a different PMT base than that used in Fig.\ \ref{fig:d2nfindiff}. 
Note that the zero is suppressed on the vertical
axis. \label{fig:happexfindiff}} \includegraphics[width=8.5cm]{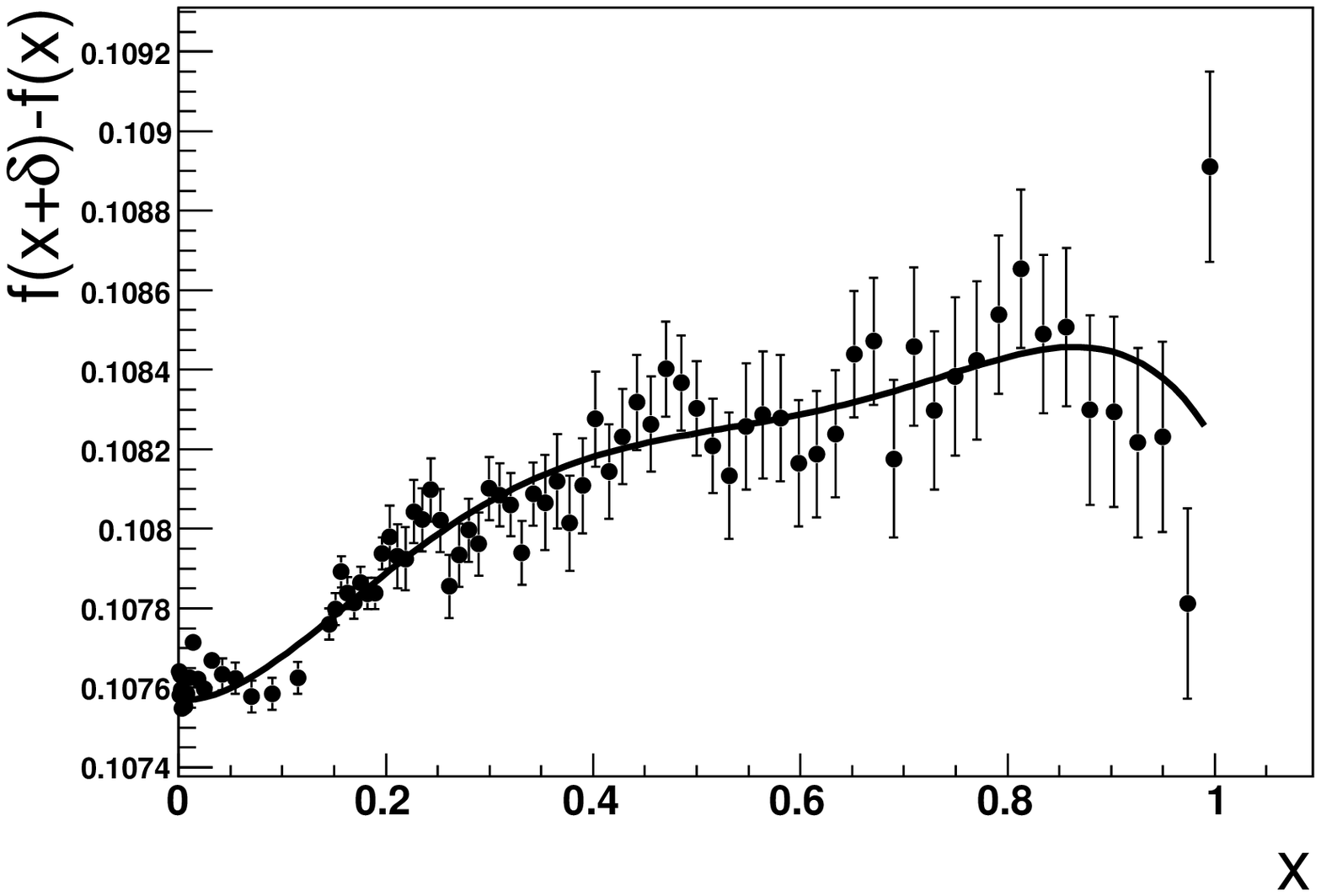} \end{figure}

Two typical PMT finite-difference response curves are shown in Figs.\
\ref{fig:d2nfindiff} and \ref{fig:happexfindiff}.  In each plot, the vertical
axis shows the difference between the total integrated signal from the PMT for a
pulse with both LEDs flashed and the signal from a pulse with just the variable
LED flashed, \(f(x+\delta)-f(x)\) from Eq.\ \eqref{eq:findif}.  The horizontal
axis shows the calculated light output from just the variable LED, \(x\), as
described below.  This
response curve is scaled by a single factor vertically and horizontally to give
a maximum of \({\sim} 1\) on the horizontal axis, to simplify fitting of
high-order polynomials.  The error on each response measurement is taken as the
RMS width of the observed distribution scaled down by the square root of the
number of events for each value of \(x\).

The curve in Fig.\ \ref{fig:d2nfindiff}, corresponding to a PMT base design
which has not been optimized for linearity, has a 16\% variation, while Fig.\
\ref{fig:happexfindiff} shows data taken on a PMT base which has been fine-tuned
to minimize non-linearity, and has only a 1\% variation in finite difference
over the range of interest.  The response function extracted from Fig.\
\ref{fig:happexfindiff} plotted against an ideal response is nearly indistinguishable 
from a 45-degree line.

To extract a response function, the PMT response curve \(f(x+\delta ) - f(x)\)
must be fit.  The PMT response for an input of size \(x\), \(f(x)\), is
approximated by an \(n^{\text{th}}\) order polynomial, \(F_n(x)\), with \(n\)
chosen arbitrarily to give an adequate parametrization of the data.  For
fitting, the initial values of \(x\) are approximated as \(x \approx f(x)\) and
the initial value of \(\delta\) is approximated as \(\delta \approx y(x{=}0)\), an
initial \(F_n(x)\) is extracted from the fit to \(y(x)\), and then the values of
\(x\) and \(\delta\) are recalculated (numerically for large values of \(n\)) by
inverting \(F_n\).  The process is iterated until the values of \(x\) and
\(\delta\) converge.  The first order coefficient of \(F_n(x)\) must be picked
arbitrarily and, in this case, was set to unity in order to equate the units of
\(x\) and \(f(x)\); the zeroth order coefficient was set to zero.

Given the small residuals of the fit, which has a \(\chi^2\) per degree of
freedom of \({\sim}1\), this setup yields a better than 1\% measurement of
the PMT response.

\section{Conclusion} 

An LED pulser has been designed to precisely determine the linearity of PMTs
using a finite-difference measurement.  The main design issue of LED cross-talk
can be eliminated with the careful design of the LED drive circuit.

\section*{Acknowledgments} This work was supported by DOE grant
DE-FG02-87ER40315. 

\bibliographystyle{plain} 
\bibliography{pulsernimbib}

\end{document}